\documentclass[10pt,a4paper]{article}
\usepackage{hep-eppsu-software}

\notoc 

\usepackage{orcidlink}

\usepackage{graphicx}
\usepackage{subcaption}
\usepackage[titletoc]{appendix}
\usepackage{xspace}

\usepackage{enumitem}
\setlist{nosep}

\usepackage[utf8]{inputenc}

\usepackage[style=numeric-comp,sorting=none]{biblatex}
\author[a]{Christina~Agapopoulou\orcidlink{0000-0002-2368-0147}}
\author[b]{Claire~Antel\orcidlink{0000-0001-9683-0890}}
\author[c]{Saptaparna~Bhattacharya\orcidlink{0000-0002-0526-6161}}
\author[d]{Steven~Gardiner\orcidlink{0000-0002-8368-5898}}
\author[d]{Krzysztof~L.~Genser\orcidlink{0000-0003-3158-8022}}
\author[e]{James~Andrew~Gooding\orcidlink{0000-0003-3353-9750}}
\author[f]{Alexander~Held\orcidlink{0000-0002-8924-5885}}
\author[g]{Michel~Hernandez~Villanueva\orcidlink{0000-0002-6322-5587}}
\author[a]{Michel~Jouvin\orcidlink{0000-0002-9501-1276}}
\author[h]{Tommaso~Lari\orcidlink{0000-0002-1388-869X}}
\author[i]{Valeriia~Lukashenko\orcidlink{0000-0002-0630-5185}}
\author[j]{Sudhir~Malik\orcidlink{0000-0002-6356-2655}}
\author[k]{Alexander~Moreno~Briceño\orcidlink{0000-0001-8415-2543}}
\author[d]{Stephen~Mrenna\orcidlink{0000-0001-8731-160X}}
\author[l]{Inês~Ochoa\orcidlink{0000-0001-6156-1790}}
\author[m]{Joseph~D.~Osborn\orcidlink{0000-0003-0697-7704}}
\author[n]{Jim~Pivarski\orcidlink{0000-0002-6649-343X}}
\author[o]{Alan~Price\orcidlink{0000-0002-0372-1060}}
\author[p]{Eduardo~Rodrigues\orcidlink{0000-0003-2846-7625}}
\author[q]{Richa~Sharma\orcidlink{0000-0002-4656-4683}}
\author[d]{Nicholas~Smith\orcidlink{0000-0002-0324-3054}}
\author[b]{Graeme~Andrew~Stewart\orcidlink{0000-0003-0182-7088}}
\author[b]{Anna~Zaborowska\orcidlink{0000-0001-6210-1921}}
\author[r]{Dirk~Zerwas\orcidlink{0000-0002-4198-3029}}
\author[s]{Maarten~van~Veghel\orcidlink{0000-0001-6178-6623}}

\affiliation[a]{Université Paris-Saclay, CNRS/IN2P3, IJCLab, Orsay, France}
\affiliation[b]{CERN, Geneva, Switzerland}
\affiliation[c]{Southern Methodist University, Dallas, Texas 75205}
\affiliation[d]{Fermi National Accelerator Laboratory, Batavia, Illinois 60510 USA}
\affiliation[e]{Technische Universität Dortmund, Dortmund, Germany}
\affiliation[f]{University of Wisconsin–Madison, Madison, WI, United States}
\affiliation[g]{Brookhaven National Laboratory,  Upton, NY 11973, United States}
\affiliation[h]{INFN Sezione di Milano, Milan, Italy}
\affiliation[i]{Physik Institute, Universität Zürich, Winterthurerstrasse 190/Building 36, 8057 Zürich}
\affiliation[j]{University of Puerto Rico, Mayaguez, Puerto Rico, USA}
\affiliation[k]{Universidad Antonio Nariño, Ibagué, Colombia}
\affiliation[l]{Laboratório de Instrumentação e Física Experimental de Partículas, Portugal}
\affiliation[m]{Brookhaven National Laboratory, United States}
\affiliation[n]{Princeton University, Princeton, NJ, United States}
\affiliation[o]{Jagiellonian University, ul. prof. Stanisława Łojasiewicza 11, 30-348 Kraków, Poland}
\affiliation[p]{University of Liverpool, Liverpool, United Kingdom}
\affiliation[q]{University of Puerto Rico, Mayaguez}
\affiliation[r]{DMLab, Deutsches Elektronen-Synchrotron DESY, CNRS/IN2P3, Hamburg, Germany}
\affiliation[s]{Nikhef, National Institute for Subatomic Physics, Amsterdam, the Netherlands}

\collaboration{ALICE}
\collaboration{ATLAS}
\collaboration{Belle II}
\collaboration{CMS}
\collaboration{DUNE}
\collaboration{ePIC}
\collaboration{LHCb}
\collaboration{MCnet}
\collaboration{WLCG}

\usepackage{lineno}  

\begin{document}


\title{The Critical Importance of Software for HEP}
\genremark{Prepared by the HEP Software Foundation, with inputs from the HEP community.}

\maketitle


\newpage

\setcounter{page}{1}

\section{Preamble}\label{preamble}

Particle physics has an ambitious and broad global experimental programme for
the coming decades. Large investments in building new facilities are already
underway or under consideration. Scaling the present processing power and data
storage needs by the foreseen increase in data rates in the next decade for
HL-LHC is not sustainable within the current
budgets~\cite{CERN-LHCC-2022-005,Software:2815292}. As a result, a more
efficient usage of computing resources is required in order to realise the
physics potential of future experiments. Software and computing are an integral
part of experimental design, trigger and data acquisition, simulation,
reconstruction, and analysis, as well as related theoretical predictions. A
significant investment in computing and software is therefore critical.

Advances in software and computing, including artificial intelligence (AI) and
machine learning (ML), will be key for solving these challenges. Making better
use of new processing hardware such as graphical processing units (GPUs) or ARM
chips is a growing trend. This forms part of a computing solution that makes
efficient use of facilities and contributes to the reduction of the
environmental footprint of HEP computing~\cite{wlcgsust}. The HEP community
already provided a roadmap for software and computing~\cite{hsfcwp} for the last
EPPSU, and this paper updates that, with a focus on the most resource critical
parts of our data processing chain.

\section{Physics Event Generators}\label{physics-event-generators}

Monte Carlo Event Generators (MCEGs) that simulate particle collisions,
or `events', to a level that allows direct comparison with
experimental data, are indispensable for the planning and analysis of
all particle physics experiments. MCEGs are vital to connect
theoretical ideas and calculations
to experimental results. They provide inputs to
detailed software models of detectors, are key components in the
evaluation of systematic uncertainties, and frequently
play an essential role in the interpretation of those measurements.

As such, the MCEGs buttress the continued success of the experimental programme
through sustained improvements responding to increasing demands on their
precision, reach, flexibility, and usability. The development, maintenance,
validation and tuning of these central assets is driven by a relatively small,
but critically important, ecosystem of researchers. The ongoing physics
exploration at the (HL-)LHC and other current experiments, and the preparation
for future facilities, requires the continuation of this strategic and vibrant
research programme. The intersectionality of MCEG developers poses serious
challenges in terms of career paths and funding opportunities, and it is
imperative that the particle physics community addresses such issues to ensure a
sustainable development of MCEGs for existing and future facilities.

As well as the MCEGs themselves, there is a critical ecosystem of tools to
exchange data with other software, HepMC3~\cite{BUCKLEY2021107310} and
LHE~\cite{ALWALL2007300} in particular; as well as data access via, e.g.,
LHAPDF~\cite{Buckley2015LHAPDF} and HEPData~\cite{Maguire_2017}. Long term,
adequate support for these infrastructure tools is vital.

\subsection{Large Hadron Collider}\label{large-hadron-collider}

The unprecedented precision in measuring the mass of the $\mathrm{W}$
boson~\cite{CMS:2024lrd} and related SM parameters at the LHC illustrates the
need for exceptional precision. With HL-LHC datasets, theoretical precision
needs to be of $\mathcal{O}(1\%)$ (or even 0.1\% for the top-quark mass). This
high precision event generation comes at the cost of computing resources, and
can reach 20\% of the total CPU budget, so considerable R\&D is needed to
ameliorate this. Significant challenges are outlined below and, although
progress has been made, there is still work to be
done~\cite{HSFPhysicsEventGeneratorWG:2020gxw,maltoni2022tf07snowmassreporttheory,10.21468/SciPostPhys.16.5.130}.

\subsubsection{Reduction of negative weights in event
generation}\label{reduction-of-negative-weights-in-event-generation}

Events with negative weights can occur in MC schemes that include
higher order (next-to-leading order, NLO; next-to-next-to-leading order, NNLO; and beyond)
corrections to cross sections and can
significantly dilute the statistical power of a sample. This statistical loss is
most significant for events that have to be propagated through the downstream steps,
which are computationally expensive. Positive resampling~\cite{Andersen:2020sjs}
and cell resampling methods~\cite{Andersen:2021mvw} have been developed by the
generator theory community -- these methods redistribute weights, without
introducing bias, thereby significantly reducing the proportion of negative
weights. Other methods include the
MC@NLO-$\Delta$-scheme~\cite{Frederix:2020trv}, an NLO-accurate matching
prescription that reduces the number of negatively weighted events. All of these
methods are currently being tested by ATLAS and CMS.

\subsubsection{Heterogeneous computing}\label{heterogeneous-computing}

Matrix element calculations, which are usually the most CPU intensive part of
higher order MCEG computations, can be offloaded to GPUs. The first benchmarking
of the gains from using the GPU version of
Madgraph~\cite{Alwall:2014hca, Hageboeck:2023blb, Valassi:2022dkc, Wettersten:2023ekm}
for leading-order calculations was
performed by the CMS Collaboration~\cite{CMS-DP-2024-086} and showed
improvement up to $\times 7$ in event generation time. The PEPPER (Portable
Engine for the Production of Parton-level Event Records)~\cite{Bothmann:2023gew}
event generator, developed by the Sherpa generator team~\cite{Sherpa:2019gpd}
parallelises the entire parton level event generation and has been tested across
a wide range of GPU architectures. The benchmarking exercises performed by
experiments are still in nascent stages, with the eventual goal that event
generators will be run on GPUs during huge MC production campaigns.

\subsubsection{Use of ML in event generation}
\label{use-of-ml-in-event-generation}

Neural networks can be used to approximate matrix elements and have been studied
in the context of loop-induced diphoton-plus-jets production through gluon
fusion~\cite{Moodie:2022flt}. Such approaches to event generation can reduce the
computation time by up to $\times 10$. ML-based hadronization models that can
replicate the performance of the Pythia 8 or Herwig generators for certain kinematic
distributions have been
developed~\cite{Ilten:2022jfm, Bierlich:2023fmh, Ghosh:2022zdz, Chan:2023ume}.
These ML models offer the possibility of being directly built from data, so can
provide insights into phenomenological models currently in use. Employing ML-based
reweighting techniques can alleviate the problem of generating MC samples
for several systematic variations -- instead of generating multiple samples,
only one sample with weights representative of each systematic variation can be
sufficient~\cite{CMS:2024jdl}. ML techniques are also being explored
for phase-space sampling and unweighting~\cite{Butter:2022rso}.

\subsection{Higgs/Top/Electroweak factories aka Higgs
Factories}\label{higgstopelectroweak-factories-aka-higgs-factories}

In addition to general-purpose generators, LEP-era generators, which are more
specialised to certain processes, have been
updated~\cite{Jadach:1999vf,CarloniCalame:2003yt,Jadach:1991by,Denner:2000bj}
and several have moved from Fortran to C++~\cite{Jadach:2022mbe, Sjostrand:2014zea}.
Modern generators, developed for Higgs factories~\cite{Kilian:2007gr, Sherpa:2024mfk}
or for LHC physics~\cite{Frixione:2021zdp,Sherpa:2019gpd},  due to their process
independent algorithms, have the capability to provide event generation not only
for the Standard Model, but also for Beyond the Standard Model (BSM). This feature
is already exploited by Belle II in the generation of samples for Dark Matter searches.
Specialised tools are necessary, e.g., for providing luminosity
spectra~\cite{Ohl:1996fi} based on Guinea Pig~\cite{Schulte:1998au}.
The interface and validation of such tools, as well as adapting the general-purpose
Monte Carlo programs to the precision of LEP era MCEGs, is crucial for the
physics programme of future Higgs factories.

The high-precision measurements at a Higgs factory poses tremendous challenges.
All processes have to be known to at least NNLO electroweak (EW) and
some, such as Bhabha scattering, even N\textsuperscript{3}LO
EW~\cite{ECFAHiggsStudy:2025}, which will require the development of new
efficient algorithms to automate the calculation of these radiative corrections.
These fixed-order corrections must also be supplemented with all-order
resummation methods. Currently, the electron parton distribution function (PDF)
at leading logarithmic (LL) and next-to-leading logarithmic (NLL) order is known, which must be matched to an exclusive description
of the photon phase space, such as a parton shower, to at least
NNLL~\cite{ECFAHiggsStudy:2025,Bertone:2019hks,Frixione:2019lga}. Such
universally applicable matching formalism between fixed-order and resummed
calculations with exclusive radiation simulations and QED showers is needed to
achieve per-mil precision for inclusive and exclusive
predictions~\cite{Frixione:2022ofv}. A complementary approach is the
Yennie-Frautschi-Suura (YFS) resummation, which was crucial to the LEP physics
programme as it was the main approach in the KK generator~\cite{Jadach:1999vf}, in which both
the order-by-order perturbative improvements and resummation of soft logarithms
is achieved in one method. This method has been implemented in process
independent fashion in Sherpa~\cite{Krauss:2022ajk}, as an example of how state-of-the-art predictions from the LEP era can benefit from technical advances in
LHC MC tools.

Specialised tools are necessary for specific processes, e.g., luminosity
determination (BHLumi~\cite{Jadach:1991by}, Babayaga~\cite{CarloniCalame:2003yt}), top threshold. Further
developments/calculations are needed to reach the required accuracy. Phase-space
sampling is being improved with adaptive multi-channel versions of VEGAS or
machine-learning methods, and can utilise GPUs.

Since the last EPPSU, the Key4hep~\cite{Ganis2022, Sailer:2020fah} software ecosystem has
emerged as a common framework supported by and supporting all Higgs factory proposals
(FCC, CLIC, ILC, CEPC), and also the Muon collider and
Electron-Ion-Collider. It provides coherent interfaces to many
generators enabling meaningful systematic comparisons of cross
sections, differential distributions and more. Furthermore, scripts to
generate events in the Key4hep ecosystem are provided that are of huge
utility to the community. As well as continued support for Key4hep, an
important goal is to develop automated generator-to-generator
comparisons for Higgs factories, i.e., complementing the tests performed
by each generator individually.

\subsection{Hyper-K and DUNE}\label{hyper-k-and-dune}

In the neutrino community, high-quality event generators are a critical need for
next-generation analyses, particularly those pursued by the accelerator neutrino
oscillation experiments Hyper-Kamiokande (Hyper-K) and the Deep Underground
Neutrino Experiment (DUNE). To make definitive measurements of neutrino and SM
properties requires unprecedented percent-level control of systematic
uncertainties, including those related to simulations of neutrino-nucleus
interactions. These interaction uncertainties, which arise to mitigate event
generator mismodeling, are particularly difficult to reduce to the required
level because of stringent theoretical demands: full final states must be
predicted over orders of magnitude in neutrino energy for multiple targets, and
the various interaction modes that contribute are each subject to complicated
nuclear effects. Delivering the simulation capability to allow Hyper-K and DUNE~\cite{DUNE_EPPSU}
to be successful will thus require advances in nuclear theory, as well as
corresponding investment in development and maintenance of the neutrino event
generators.

Although neutrino event generators are at an earlier stage of development and
supported by a smaller workforce, there are many challenges shared with
simulation efforts in collider experiments and other branches of
HEP~\cite{campbell2024eg}. Specific obstacles to the neutrino community must also be
overcome. The field is heavily reliant upon nuclear research, which is often
funded for other applications. Technical barriers already solved for the LHC
still remain an issue for the neutrino community due to a lack of dedicated
effort, e.g., there is no common event format (although one has 
been proposed~\cite{gardiner2024nuhepmc}) and no standardised interfaces for related
software. The implementation of new models also typically remains very labour
intensive, with adaptation of LHC techniques for streamlining the work only just
beginning~\cite{PhysRevD.105.096006}. Supporting the
wide variety of new physics searches envisioned for future experiments, will, in
particular, require novel approaches.

\subsection{Recommendations}\label{recommendations}

In view of the current state and importance of MC generators (and their
growing projected importance for the HL-LHC) and the lack of resources
routinely plaguing the generator theory community, the following
recommendations are made:

\begin{enumerate}
\def\labelenumi{\arabic{enumi}.}
\item
  Generator theory community funding, including fellowships for students, and
  ensuring expert continuity by rewarding work on computational aspects of
  generators with secure positions.
\item
  Supporting workshops on specific aspects of MC production that help bridge the
  gap between theory and experiment and provide travel awards for early career
  scientists.
\item
  Support for developments of Higgs factory event generators and the
  Key4hep ecosystem to bring together Monte Carlo generators and
  experiments into a coherent framework.
\item
  Ensure support for the neutrino event generators, with their specific
  needs, and encourage as much sharing as possible with the other
  communities.
\item 
  Provide sufficient long term support for MCEG infrastructure codes.
\end{enumerate}

\section{Detector Simulation}\label{detector-simulation}

As the volume of data collected by high energy experiments increases, so
does the need for large, accurate Monte Carlo (MC) datasets. The simulation
of the particle propagation in the detectors, and the response of readout
channels, is a major component of experiment computing budgets.

\subsection{Baseline Simulation Code
Needs}\label{baseline-simulation-code-needs}

Almost every HEP experiment uses the Geant4~\cite{ALLISON2016187, 1610988, GEANT4:2002zbu}
simulation toolkit. Its
development and maintenance over the long timescales, driven by the needs of
experiments using the toolkit, are of critical importance and need to be
supported. Higher intensity and luminosity require more precision, as does the
use of higher granularity calorimeters. The main ongoing
developments~\cite{g4inputs} are the implementation of more detailed EM models
and rarer processes, the inclusion of low energy nuclear physics effects, and
the continued tuning of string models. The
FLUKA.CERN~\cite{FLUKA:new_capabiliies, battistoni_2024_stv3r-7ar12} hadronic
physics code have been made available as an option in Geant4, and use of MC
generator code for hadron interaction and decay processes is being explored.
The validation of Geant4 physics against thin and thick target experimental data
for each release is essential. FLUKA.CERN itself, which is being modernised to
C++, remains critical for radiation transport and shielding calculations.

The computing load from electronics simulation and pileup overlay in the
digitisation step is heavily dependent on detector design and consumes
significant computing for all LHC experiments. Re-use of background
induced digitised output has been deployed by all of them.

Since the last EPPSU~\cite{European:2720131}, most LHC experiments have transitioned
their simulation applications to multithreaded execution, supported by Geant4,
and modernised their code. ALICE uses sub-event parallelism, which is now being
implemented in Geant4. ATLAS and CMS have
reported~\cite{ATLAS:2024_swcomp_run3, CMS:computing_results_2023} throughput
increases of a factor 1.5 to 2 at negligible cost in physics accuracy since the
beginning of Run 2, from improvements in their own code and in Geant4. This is a
process that can, and must, continue.

\subsection{Novel Computing
Architectures}\label{novel-computing-architectures}

Market trends favour computing using GPUs, which can be more energy efficient if
used intensively, in addition to porting to more power efficient CPU
architectures, such as ARM. There is thus considerable interest in running the
compute intensive simulation on GPU-CPU hybrid architectures. Efforts are
focusing on the calculation of EM processes on GPUs, which can be up to two
thirds of the run time in HEP collider detector simulations. The AdePT
\cite{Amadio_2023, AdePT} and Celeritas~\cite{osti_2361202} projects, working with the
experiments, have demonstrated the possibility to perform the EM part of
simulations in simplified setups on GPUs faster than on
CPUs~\cite{CHEP2024:celeritas_improvements, CHEP2024:adept_gpu_em_transport},
while simulating hadronic physics and
processing experiment-specific energy deposits on CPUs. Bottlenecks,
particularly in the geometry, have been identified that are being addressed by a
new surface geometry approach. Further work is needed to completely integrate GPU-based
simulation modules into experiment software frameworks, but the goal of using
GPUs for MC datasets for the major LHC experiments by the start of Run 4 appears
within reach. The other leading candidate for GPU simulation is the simulation
of optical photons, which is a computing bottleneck for LHCb and several
Intensity Frontier and Dark Matter experiments~\cite{opticksCHEP}.

\subsection{Fast Simulation}\label{fast-simulation}

Since the last Strategy Update~\cite{European:2720131}, fast simulation
techniques have been developed and used in the experiments' MC
production. Parametrised and generative machine learning (ML) approaches for
calorimeter shower simulations are used (e.g., \cite{af3}), or planned to be
used, in a substantial fraction of MC production by ATLAS, CMS, and
LHCb. Generative adversarial networks for background are
already developed and planned to be used in Belle II. The use of fast simulation
is expected to increase in the next few years,
with more applications transitioning from R\&D to production. More sophisticated
ML methods, e.g., INN~\cite{kim2021innmethodidentifyingcleanannotated},
CFM~\cite{tong2024improvinggeneralizingflowbasedgenerative} are being explored.
Fast simulation R\&D includes the generation of low-level detector output,
tracking with simplified geometry, fast inner detector simulation based on
parametrised models (e.g., FATRAS used by ATLAS~\cite{fatras}) or using
ML-methods, and the correction of fast simulation output to match detailed
simulation training samples. ATLAS's FastChain tool combines fast tracking and
fast calorimeter simulation in one workflow. CMS, LHCb, and ALICE are
investigating the more radical approach of generating high-level outputs for
physics analysis based on the event generators' output, saving greatly on
storage. Since fast simulation techniques still need large, accurate simulation
samples for training, they do not remove the need for detailed particle tracking
simulations, but they can reduce the total simulation time and the size of the
simulated samples.

Large experiments have dedicated fast simulation development teams, an approach
that can obtain the best performance for a specific application, but smaller
experiments cannot afford this level of specialisation. Public datasets for the
training of fast simulation models have been provided by the experiments, and an
experiment-independent framework to develop ML-based fast calorimeter simulation
models has been developed in Geant4. The field is in rapid development and the
balance between experiment-specific code and the common code still needs to be
found, but it is clear that the development of common frameworks and tools needs
to be supported to make the best use of a limited pool of developers and avoid
unnecessary duplication of efforts.

Fast simulation and ML-based approaches to mitigate the resource usage
of the current pileup mixing approaches are being considered by ATLAS
and CMS.

\subsection{Future Experiments}\label{future-experiments}

The simulation requirements for the design of future detectors are very
different from those of the running experiments. While large samples are not
usually required, it is important to be able to simulate many
different design variations, so flexibility and ease of use are
essential. Studies for detectors at proposed future colliders are using
the common Key4hep~\cite{Ganis2022} framework. Gaseous detector R\&D relies
heavily on Garfield++\cite{garfield++} for detailed modelling of signal
formation. Long term support of these tools is critical.

The FCC-ee experiments should collect datasets similar or
larger in size than that of HL-LHC experiments~\cite{FCCee-Resources}. They will perform precision
measurements in a cleaner environment than hadron colliders, so simulation
needs to be at least as fast, but also more accurate than for the HL-LHC. The
Muon Collider must handle a large background from beam muon decays, which
requires beam background simulation to be overlaid with simulated collision
events. FCC-hh experiments will need the extension of physics models used in
hadronic interaction simulation to higher energy, as well as overlay of up to
1000 pileup events per signal event and thus huge event, and sample, sizes.

\subsection{Recommendations}\label{recommendations-1}

\begin{enumerate}
\def\labelenumi{\arabic{enumi}.}
\item
  Improve the level of support for critical core simulation software code in
  Geant4, improving the fidelity and speed of physics models, and ensuring a
  generational transition before current experts retire as well.
\item
  Provide suitable long-term funding to maintain new developments which
  come to fruition.
\item
  Invest in R\&D lines that can explore simulation on accelerated
  hardware, and move into production if successful.
\item
  Invest in fast simulation support, particularly using ML, where this
  can be adapted to and support different experiments and detector
  studies, as well as benefit current experiments through exploring
  novel approaches.
\end{enumerate}

\section{Reconstruction and Software
Triggers}\label{reconstruction-and-software-triggers}

LHCb~\cite{LHCb:RTDP} and CMS~\cite{CMS:Detector_R3}
are now successfully operating GPU farms for their
Run 3 software trigger systems and ALICE~\cite{ALICE:LS2_upgrades} uses such a farm for
synchronous reconstruction,
online calibration of detectors and compression of the real data. This
effort has produced impressive, sustainable production software providing
speed-ups, portability and vendor
independence~\cite{Concas:2024cyu, Aaij2020Allen,MathesP3MA2017}. The
upgrade has enabled ALICE and LHCb to fully adopt real-time analysis strategies,
discarding raw data in favour of highly compressed or reconstructed
data. This reduces storage needs and meets baseline physics goals, demonstrating that
online calibration and detector alignment in quasi real-time is possible.
ATLAS~\cite{ATLAS:trigger_LS3} and CMS~\cite{CMS:enriching_phys_programs}
have increased their bandwidth allocation for real-time
analysis triggers in Run 3 by factors of 2 to 3, allowing them to
collect up to an order of magnitude more physics events beyond
baseline for low mass exotic particle resonance searches.

The need for faster reconstruction and data processing algorithms is being
widely addressed by ML and AI algorithms. AI tools are already used extensively
in HEP offline software and are increasingly found in online software as well as
hardware DAQ systems. For example, ATLAS uses, and continuously improves, transformer 
neural network based heavy flavour jet and tau identification in its high-level
trigger (HLT)~\cite{ATLAS:perf_commissioning} in Run 3.
CMS deployed a variational autoencoder based on
an anomaly detection demonstrator in their Level-1 trigger in 2023~\cite{CMS-DP-2023-079},
making use of HLS4ML~\cite{fastml_team_hls4ml}, a HEP community-driven
software package for high-level synthesis of ML algorithms for FPGAs. In Run 3,
LHCb has deployed monotonic Lipschitz networks at the HLT
stage~\cite{LHCb:Lipschitz} to select heavy-flavour-quark decays in an efficient
and robust way, as well as adding neural network models to their HLT to
improve track purity~\cite{neuralmodelHLT1LHCb}.

Many experiments have recognised the compute intensive need for tracking
capabilities in the trigger for improved event selection. Since Run 3, LHCb
reconstructs tracks on GPUs at the LHC bunch crossing rate, identifying quality
particle candidates for offline analysis for every event~\cite{LHCb:HLT}. This
avoided hardware-based first level trigger selections impacting its
physics programme, such as measurements for CP violations in heavy flavoured
hadron decays. Also from Run 3, CMS runs track reconstruction on GPUs based on the
pixel detector data and will make a big advance at the HL-LHC
by running tracking for particle flow reconstruction in a hardware-based
first level trigger. The next important advancement will be 4D reconstruction:
track reconstruction with the use of hit time information. This will be
applicable at the HL-LHC and to future hadron, electron-positron, and deep
inelastic scattering collider facilities for background mitigation. The tracking
R\&D area has fostered the A Common Tracking Software (ACTS) project~\cite{Ai2022Common},
an excellent example of a common software toolkit used by several experiments to
tackle tracking challenges as a community. It is already proving a useful
platform for common tools, such as traccc~\cite{yeo_2023_8119769, traccc}, a
demonstrator project for GPU-based tracking.

However, the rapidly evolving software landscape in HEP is highlighting a
struggle to keep physicists sufficiently trained to contribute to reconstruction
and trigger software in a useful way (e.g., C++ expertise is increasingly hard to find). 
There is a growing need for a dedicated group of software and
physics professionals working together. Proper software engineering leads to
immense benefits and is a foundation for software sustainability, as well as
efficient and reliable, hence more environmentally sustainable, computing.

The community has identified several key themes in reconstruction and software
triggers that need to be intensively explored in the next decade to achieve our
physics goals. Not all areas are relevant for every experiment, but combined
efforts in the proposed fields will broadly benefit future HEP programmes.

\subsection{Enhanced heterogeneous computing
software}\label{enhanced-heterogeneous-computing-software}

Utilising heterogeneous architectures, such as GPUs and FPGAs, appears
inevitable in several different areas, yet one of the main challenges is
anticipating the technological landscape in 5-10 years. While GPUs are one of
the most promising accelerators, alternative accelerator and processor
technologies, such as RISC-V, IPUs and AI-processors, are evolving rapidly and
need to be closely monitored. The community should aim to improve the
flexibility of its software in order to explore and adapt to a variety of
architectures, without being tied to a specific hardware vendor.

\subsection{Real-time analysis}\label{real-time-analysis}

Beginning with Run 3 ALICE has been operating triggerless continuous readout,
and future online data acquisition software in HEP and NP trends towards operating
triggerless or streaming readouts, to avoid online event selections that would
limit primary physics goals. As rates increase, these techniques move closer
to the detector, hosted on less flexible systems, which requires very careful
system design, as well as operational models and validation. Selective
persistency models will likely need to be further exploited to reduce storage
pressure, along with tools for real-time detector calibration.

\subsection{Integration and support of AI/ML
algorithms}\label{integration-and-support-of-aiml-algorithms}

Interest in AI/ML solutions is rapidly increasing as they lend themselves well
to low latency approaches for real-time analysis and fast data processing.
Reaching the latency and throughput requirements at the lowest levels requires
specialised expertise, posing a significant challenge and open source
collaboration projects should be pursued further. For example, the HLS4ML
project~\cite{fastml_hls4ml,Duarte:2018ite} has provided an
easier, faster means of developing ML algorithms for FPGAs, together with the
capability to generate bitwise correct C++ representations for simulation,
thus reducing the gap in firmware expertise.

\subsection{Exploitation of precise timing in the reconstruction
sequences}\label{exploitation-of-precise-timing-in-the-reconstruction-sequences}

Many HEP experiments are planning, or incorporating, precise timing
($\mathcal{O}(\mathrm{ps})$) detectors, in order to maintain performance in
ultra-high density environments (e.g., at Belle-II and HL-LHC). For the HL-LHC
and future colliders, reconstruction techniques that use hit level timing
(4D reconstruction) can much more efficiently distinguish signal from
background. Adapting reconstruction algorithms to keep pace with evolving sensor
and detector R\&D is a significant challenge and requires community effort.

\subsection{Enhanced software maintenance and Quality Assurance
(QA)}\label{enhanced-software-maintenance-and-quality-assurance-qa}

Maintaining validated online and offline reconstruction software frameworks is a
significant challenge, especially with the move towards heterogeneous systems.
Dedicated funding is essential to achieve this. Proper software engineering and
common tools for continuous integration and automation of quality assurance
tasks will ultimately conserve resources, as will identifying key software
developments that are widely applicable. Common software can be more robustly
tested and validated, with the benefits shared by many. This will foster a
sustainable software environment and at the same time increase computing
efficiency to improve physics reach.

\subsection{Recommendations}\label{recommendations-2}

\begin{enumerate}
\def\labelenumi{\arabic{enumi}.}
\item
  Support R\&D in heterogeneous architectures along with the associated
  skill development.
\item
  Invest in further development of real-time and streaming readout systems with
  emphasis on reliability, to address growing storage pressures in future
  ambitious HEP programmes.
\item
  Support ML/AI developments, in particular platforms for sharing ML/AI
  methods and tools specific to HEP applications and ML/AI
  implementations on novel devices.
\item
  Invest in new algorithm developments using evolving precision timing
  detector technology to significantly advance particle reconstruction
  in dense environments.
\item
  Support common software solutions that encourage collaboration between
  experiments.
\item
  Invest in sustainable software development by promoting software
  maintenance, validation, quality assurance, dedicated software groups
  and developer training.
\end{enumerate}

\section{Data Analysis}\label{data-analysis}

The data analysis software ecosystem, that is, the common frameworks, toolkits
and packages typically employed in HEP data analysis, has generally migrated
from pure C++ to a C++/Python hybrid over the last 10 years. The increased
prevalence of Python has been driven by usability benefits and changes in
language preferences of physicists, with community initiatives such as
Scikit-HEP~\cite{Rodrigues_2020} promoting and consolidating Pythonic HEP
analysis tools and the ``Python in HEP'' Developer's Workshop
(PyHEP.dev)~\cite{alshehri2024pyhepdev2024workshopsummary} bringing together
experts to form new collaborations. Typically, earlier stages of data analysis
use C++ and later stages use Python as a ``glue language'' to steer frameworks
such as ROOT. Further emphasis on interoperability and sustainability have
encouraged many tools to implement/enhance their Python compatibility, e.g.,
through C++-Python wrappers, such as ROOT's Python Interface. To overcome
performance limitations, many Python packages have adopted an \emph{array
programming} interface, where large arrays of data are passed into a compiled
layer for performance, focusing the Python portion of codebases to the task of
dispatch and metadata manipulation. A recent addition to the community is a
small but growing group of analysts employing Julia. Whilst not expected to
become the mainstream in the immediate future, the significant performance
benefits against Python and usability benefits against C++ make Julia an
appealing language.

In parallel, data analysis frameworks are improving integration with computing
infrastructure. The ROOT framework has developed the
RDataFrame~\cite{ROOT:RDataFrame} interface, with improved
multithreading and interfaces to Spark~\cite{10.1145/2934664}
and Dask~\cite{matthew_rocklin-proc-scipy-2015}, two popular distributed
computing technologies. In the Scikit-HEP ecosystem,
awkward-array~\cite{Pivarski_2020} and Coffea~\cite{Smith_2020} have closely
integrated Dask to achieve scalability.

Given the wide range of tools available to HEP analysts, interoperability across
the software ecosystem is necessary, with the need to build well-specified and
interoperable data structures for all stages of analysis. The ROOT RNTuple data
serialisation interface, replacing the TTree interface, has a clear
specification as well as significant enhancements in support for multithreading
and asynchronous I/O~\cite{ROOT:RDataFrame_prod_path}. The
HEP Statistics Serialisation Standard (HS3) aims to provide a common, framework
agnostic serialised description of the components involved in fitting
statistical models (models, datasets, likelihoods,
etc.)~\cite{HS3github}.

\subsection{Machine learning and automatic
differentiation}\label{machine-learning-and-automatic-differentiation}

ML has had a transformative impact on HEP and become a core part of HEP data
analysis. Traditionally, ML algorithms such as decision trees and neural
networks (NNs), are used to efficiently classify events. More recently, the use
of ML in analysis has extended to statistical inference, with results from
ATLAS~\cite{ATLAS:2024_hzz_higgs, ATLAS:neural_sim_inference}
using simulation-based inference, wherein ML classifiers are trained on
simulation and used to construct statistics, such as likelihood ratios, which
can be evaluated on data. We expect ML to keep playing a significant role in the
future.

Automatic differentiation (AD), techniques for the fast and precise computation
of derivatives of functions, has potential to provide significant performance
improvements across HEP analysis. In statistical inference and fitting, where
derivatives of functions are evaluated many times, AD has been shown to be
significantly faster than existing methods, e.g., in the codegen backend in
RooFit~\cite{Hageboeck:2020dyv}. AD also enables applications such as
systematics-aware training of NNs.

\subsection{Sustainable software development and
analysis}\label{sustainable-software-development-and-analysis}

Sustainability has become a major focus of data analysis software and is
typically discussed in the context of environmental and economic impact,
software lifecycles and analysis reproducibility.

Many important software packages in HEP analysis are developed and
maintained by early career researchers, who are typically employed on
fixed-term contracts of no more than a few years and for whom the time
working on software is poorly recognised. Without direct support for,
and recognition of, this work, packages can become insufficiently
maintained to be useful to the community, with very negative
impacts upon analysis, and, by implication, on the experiments' physics
programmes and the future usability and preservation of results.

Analysts are increasingly encouraged to make analyses and all related artefacts
findable, accessible, interoperable and reusable
(FAIR)~\cite{wilkinson_fair_2016, Chen_2022, Duarte:2022job, FAIR4AIWorkshop}. The
expertise and tooling required has developed rapidly over recent years. In
particular, developments in the domain of workflow management
systems~\cite{cwl,snakemake,luigilaw,yadage} have enabled analysts to construct
analyses in the form of structured workflows. The REANA reusable analysis
platform~\cite{REANA} extends this to provide a data analysis platform from
which to develop reusable analyses. Increased functionality and user uptake of
CI/CD systems provides tooling to ensure the integrity and reliability of
analyses and software alike.

Reinterpretation of results is helped greatly by this approach and is itself
supported by important tools such as Rivet~\cite{10.21468/SciPostPhys.8.2.026}.

A topic of increasing interest for HEP is the importance of open data
(for research and education), and data preservation. This requires
effort in all aspects of software, but especially analysis, and does
not necessarily have a clear end date, even outlasting the experiments which
created the data.

\subsection{Analysis facilities and data analysis at
scale}\label{analysis-facilities-and-data-analysis-at-scale}

With the increasing scale of data to be processed in HEP analyses, it is
vital that analysts can employ large scale distributed computing resources.
Analysis facilities (AFs) provide an approach to analysis infrastructure in
which analysts can interface with these resources whilst also being able to
prototype and execute their analyses interactively~\cite{WLCG:ana_fac_wp}. Demonstrators
such as the Analysis Grand Challenge
initiative~\cite{Held:2022RC}, in which physics analyses
are performed to test the technologies which will be required for the HL-LHC,
and many prototype AFs have shown promise in this model.

\subsection{Recommendations}\label{recommendations-3}

\begin{enumerate}
\def\labelenumi{\arabic{enumi}.}
\item
  Support the continued development of analysis tools, strongly
  emphasising interoperability and standards, while also exploring new
  R\&D lines.
\item
  Ensure recognition and secure positions for the development of
  analysis software and supporting technical contributions that
  result in increased usability, efficiency and sustainability.
\item
  Improve training and infrastructure to make our analyses FAIR.
\end{enumerate}

\section{Training and Careers}\label{training-and-careers}

\subsection{Current Status}\label{current-status}

Present day HEP experimental collaborations have hundreds or thousands of users.
They have internal structures and onboarding processes to help new users learn
and jumpstart software and physics analysis contributions. While most analysis
tools and software frameworks taught are experiment-specific, a considerable
amount of training is invested in common software tools and skills like Python,
C++, ML, CI/CD, and analysis preservation and reproducibility, etc. There is a
growing consensus in the HEP community on the need for common training programmes
to bring researchers up to date with new software
technologies~\cite{HSF-CWP-2017-02, Snowmass:2021_community_engagement_frontier, Snowmass:2021_CEF_report, Snowmass:2021_career_pipeline}.
As a result, in the past 5 years the HSF~\cite{HSFTraining}, together with Institute for
Research and Innovation in Software for High Energy Physics (IRIS-HEP)~\cite{IRISHEP:Training},
CERN EP-SFT group and FIRST-HEP~\cite{FIRST-HEP}, and partnering with The
Carpentries~\cite{TheCarpentries}, and PyHEP~\cite{HSFPyHEP} have developed material for an
introductory, intermediate and advanced HEP software curriculum~\cite{HSFTrainingCenter} taught
to HEP users. In addition, several of our major conferences have tracks on
software training and corresponding efforts. Other specific training efforts
include CoDAS-HEP~\cite{CODAS-HEP}, Bertinoro~\cite{INFN:ESC_school}, CERN School of
Computing~\cite{CERN:computing_school} and CERN OpenLab Software
workshops~\cite{CERN:SW_workshops}. Thus far,
over 3000 users in HEP, related Nuclear Physics and computing areas have been
trained. Current training work has also been published in journals and
conference proceedings~\cite{Malik:2919564, Malik:chep2023_training_outreach, Malik2021Software, 10.3389/fdata.2025.1497622}.

\subsection{Challenges and
Opportunities}\label{challenges-and-opportunities}

While a lot of progress has been made with respect to finding and establishing
common training across HEP, several challenges still remain. Currently, each
experiment may have its own training programmes that are specific to its
needs, but for HEP wide effort, committed funding sources for HEP have been
lacking. A central organisation like the HSF is needed to facilitate cooperation
and engage people in common training efforts. Incentives and recognition are
essential to be able to recruit mentors and tutors who invest time in training.
Scalability and sustainability are key to achieve a steady state, where the
scale of training activities can match the number of incoming students each year.
New opportunities, such as using Large Language Models for training, need to be investigated.
Training scope and curricula must evolve to meet the needs of the community.
New instructors and materials developers recruited and trained continuously are required to
sustain a long-term effort. 
Commitments from major HEP labs and experiments to workforce training bring significant value. 
Diversity
and inclusion should be a guiding principle to develop the workforce pipeline
and should engage the broader community that it represents.

The benefits of investing in training like this are considerable, directly for
HEP and for supporting a set of broad interdisciplinary skills that help
research software through, e.g., national and institutional research software
engineer groups.

\subsection{Career Support and
Recognition}\label{career-support-and-recognition}

For successful HEP computing, training provides a crucial step to recruit our
future workforce and provide wide societal benefits. Most training is led by the
early career researchers who are trying to build a career. The current mindset
favours visibility and recognition of those working on data analysis projects or
detector hardware rather than those invested in software and computing. This
must change if we want to meet future challenges over decades and in multiple
experiments, which increasingly requires large and long-lived software projects.
We must provide such opportunities and provide a career path for those
involved in software and computing, including in training, by creating job
opportunities at labs and universities on par with those involved in detector
construction~\cite{hsfcwp}. This must be supported by increased visibility like publications,
conferences opportunities and recognition of these contributions as valid
scientific accomplishments.

\subsection{Recommendations}\label{recommendations-4}

\begin{enumerate}
\def\labelenumi{\arabic{enumi}.}
\item
  Create training-the-workforce funding opportunities to engage
  students, postdocs who can be mentors and tutors for the
  training programmes.
\item
  Align software training efforts with other research software bodies to
  maximise impact.
\item
  Increase job opportunities for those investing significantly in
  training and computing.
\end{enumerate}

\section{Conclusion}\label{conclusion}

This paper describes the software and computing challenges that are
critical for the success of HEP experiments. The main motivation for
addressing them is the physics output of current and future experiments
with their increased data volume and need for precision. However, this
also responds to the requirement to make our computing environmentally
sustainable. The main way to address these needs is by improving
software efficiency. As many aspects of these challenges are common to
several, if not all, experiments, global collaboration
around software, as supported by the HSF, has been extremely valuable to the
work done in the last 10 years. Continuing the work that the HSF undertakes 
requires a commitment from labs, universities and funding agencies.

To address all these challenges, the main asset is \emph{skilled people}
who can make a \emph{viable career} in our field. Addressing this
challenge, to recruit, train and retain scientific software developers,
is the key message for the future.

\section{Acknowledgements}

The editors would like to thank the many people from the HEP community who provided input to this document and made possible this summary on the software and computing challenges faced by HEP experiments. We also thank the many projects and organisations that have provided funding to several colleagues, making possible their contributions. In particular the U.S. National Science Foundation (NSF) Cooperative Agreement PHY-2323298 (IRIS-HEP); the OpenMAPP project, via National Science Centre, Poland under CHIST-ERA programme (NCN 2022/04/Y/ST2/00186); Brookhaven National Laboratory, a U.S. Department of Energy, Office of Science facility; Fermi National Accelerator Laboratory (Fermilab), a U.S. Department of Energy, Office of Science, Office of High Energy Physics HEP User Facility (Fermilab is managed by FermiForward Discovery Group, LLC, acting under Contract No. 89243024CSC000002).

\newpage
\sloppy
\printbibliography

\end{document}